\documentclass[seceq,supplement]{ptptex}
\usepackage{wrapft}

\usepackage{graphicx}

\def\fsl#1{\setbox0=\hbox{$#1$}           % set a box for #1 
   \dimen0=\wd0                                 % and get its size
   \setbox1=\hbox{/} \dimen1=\wd1               % get size of /
   \ifdim\dimen0>\dimen1                        % #1 is bigger
      \rlap{\hbox to \dimen0{\hfil/\hfil}}      % so center / in box
      #1                                        % and print #1
   \else                                        % / is bigger
      \rlap{\hbox to \dimen1{\hfil$#1$\hfil}}   % so center #1
      /                                         % and print /
   \fi}          
\title{%        %You can use \\ for explicit line-break
Composite Avenue beyond the Standard Model
\subtitle{Legacy of Sakata in LHC Era}    %Use this when you want a subtitle.

\author{%       %Use \scshape  for the family name
Koichi \textsc{Yamawaki}%$^{1,}$
}

\inst{%         %Affiliation, neglected when [addenda] or [errata]
Kobayashi-Maskawa Institute for the Origin of Particles and the Universe (KMI), Nagoya University, Nagoya 464-8602, Japan
}

}

\abst{
Higgs boson may be a composite particle as Sakata vigorously looked for never-ending substructures of Nature. He proposed the Sakata model for hadrons, which was the prototype of the quark model and thus
lauched the last Revolution in particle physics continued all the way up to Kabayashi-Maskawa work which completed  the Standard Model today. Inspired by the Sakata's spirit 
 we shall discuss composite Higgs boson in various models of our own for the dynamical symmetry breaking with
large anomalous dimension: The techni-dilaton  in the walking technicolor (WTC)   with $\gamma_m \simeq 1$, the $\bar t t$ composite (``top-Higgs'') in the top-quark condensate model with $\gamma_m \simeq 2$, and their variants in the models with $1<\gamma_m <2$ (strong ETC Technicolor, etc.). Among others we will focus on WTC which has an approximate  scale symmetry in the region relevant to the dynamical mass generation. 
Such a conformal gauge dynamics is characterized by the essential singularity scaling, breakdown of the Ginzburg-Landau/Gell-Mann-Levy effective theory, and also by a large anomalous dimension $\gamma_m =1$.  In contrast to the folklore that Technicolor is a ``Higgsless theory'', there exists a composite Higgs, techni-dilaton,  in the WTC as a composite pseudo Nambu-Goldstone boson associated with the spontaneously broken (approximate) scale symmetry, with its mass  
only arising from the (nonperturbative) scale anomaly and hence being much smaller than those of other techni-hadrons. The techni-dilaton has a mass typically of order $500-600 \, {\rm GeV}$ and can  be discovered at LHC.
 We shall also touch upon the endeavor to discover WTC on the lattice.
}

\begin{document}

\newcommand{\beq}{\begin{equation}}
\newcommand{\eeq}{\end{equation}}
\newcommand{\beqs}{\begin{eqnarray}}
\newcommand{\eeqs}{\end{eqnarray}}
\newcommand{\zm}{z_{m}}
\newcommand{\ep}{\epsilon}
\newcommand{\tr}{\textrm{Tr}\,}
\newcommand{\condense}{\langle \bar{T}T \rangle}
\newcommand{\Tr}{\hbox{Tr}}
\newcommand{\Ln}{\hbox{Ln}}
\newcommand{\FT}[1]{{\cal FT}#1}
\newcommand{\hyper}[4]{ \hbox{F}({#1},{#2},{#3};{#4}) }  
\newcommand{\half}{{1\over2}}
\newcommand{\VEV}[1]{\langle #1 \rangle}
\newcommand{\bra}[1]{\langle #1 |}
\newcommand{\ket}[1]{| #1 \rangle}
\newcommand{\xSB}{\mbox{$\chi SB$}}
\newcommand{\DxSB}{\mbox{$D\chi SB$}}
\newcommand{\Lag}{\mbox{${\cal L}$}}
\newcommand{\DISP}{\displaystyle}
\newcommand{\intp}[1]{\int {d^4 #1 \over (2\pi)^4}}
\newcommand{\integ}[2]{\int_{#1}^{#2}\!\!}
\newcommand{\integg}[2]{\int_{#1}^{#2}}
\newcommand{\TeV}{\hbox{TeV}}
\newcommand{\GeV}{\hbox{GeV}}
\newcommand{\MeV}{\hbox{MeV}}
\newcommand{\xbar}[1]{#1 \hspace{-5.5pt}/}
\newcommand{\unit}{{\bf 1}}
\newcommand{\SxSB}{\mbox{$S\chi SB$}}
\newcommand{\NJLd}{\mbox{$\hbox{NJL}_{D<4}$}}
\newcommand{\EQ}[1]{(\ref{#1})}
\newcommand{\GB}{ \langle \alpha G_{\mu \nu}^2 \rangle}
\newcommand{\siml}{\hspace{0.3em}\raisebox{0.4ex}{$<$}
\hspace{-0.75em}\raisebox{-.7ex}{$\sim$}\hspace{0.3em}}
\newcommand{\simg}{\hspace{0.3em}\raisebox{0.4ex}{$>$}
\hspace{-0.75em}\raisebox{-.7ex}{$\sim$}\hspace{0.3em}}

\renewcommand{\thefootnote}{\#\arabic{footnote}}
\def\fsl#1{\setbox0=\hbox{$#1$}           % set a box for #1 
   \dimen0=\wd0                                 % and get its size
   \setbox1=\hbox{/} \dimen1=\wd1               % get size of /
   \ifdim\dimen0>\dimen1                        % #1 is bigger
      \rlap{\hbox to \dimen0{\hfil/\hfil}}      % so center / in box
      #1                                        % and print #1
   \else                                        % / is bigger
      \rlap{\hbox to \dimen1{\hfil$#1$\hfil}}   % so center #1
      /                                         % and print /
   \fi}       
\maketitle

\section{Introduction}

Composite idea has been a main stream of modern physics, from atoms to nuclei, hadrons  and eventually to quarks. The latest 
establishment of this idea, the quark, proposed by 
Gell-Mann~\cite{GellMann:1964nj} and Zweig~\cite{Zweig:1964zz} 
 in 1964 had its precedence, the Sakata model~\cite{Sakata:1956hs}, which was  born in Nagoya back in 1955 (published in 1956).  Here is the phrase in Ref.{\cite{GellMann:1964nj} :  ``A formal mathematical model based on field theory can be built  up for the quarks 
exactly as for $p,n,\Lambda$ in the old Sakata model, $\cdots$ For the weak current, we can take over  from the Sakata model the form suggested by Gell-Mann and L\'evy, $\cdots$''. The influence of the Sakata model is more eminent in the ace model ~\cite{Zweig:1964zz}  which consists of  the fundamental triplet ($p_0, n_0,\Lambda_0$) mocking up the Sakata model triplet ($p,n,\Lambda$). In 2006 we had a symposium at Nagoya celebrating Jubilee of the Sakata model. The poster of the symposium~\cite{Sakata50} was designed to describe the development from $(p,n,\Lambda)$ into  ($u,d, s$), by moving  $(p,n)$ (and $\Lambda$)  gradually upside-down to $(u,d)$ (and $s$).  
At  Nagoya the Sakata model developed into the ``Nagoya model''~\cite{Maki:1960ut}}   based on the correspondence of fundamental triplet ($p,n,\Lambda$) to the lepton triplet ($\nu,e,\mu$).  After discovery of $\mu$-neutrino,
the model was modified into the Maki-Nakagawa-Sakata model~\cite{Maki:1962mu} which proposed the neutrino mixing $\nu_e - \nu_\mu$ (so-called MNS matrix) and also a new entry $p^\prime$ into the fundamental 
triplet  so as to be extended into the fundamental quartet $(p,n,\Lambda,p^\prime) $ corresponding to the quartet of leptons $(\nu_e,e,\mu,\nu_\mu)$. This quartet model actually turned out to be  a  stepping stone 
for Maskawa and  Kobayashi, both the Sakata's disciples at Nagoya University, to jump into  the Kobayashi-Maskawa model~\cite{Kobayashi:1973fv}  proposed in 1973. Thus the Nagoya University  tradition of composite  model  launched and finalized the latest Revolution, the Standard Model (SM) Revolution.  Now in the occasion of the Sakata Centennial,  I am going to talk about composite Higgs model developed by our group at Nagoya in recent 30 years.

The Origin of Mass is the most urgent issue of the particle physics today and is 
to be resolved at the LHC experiments. 
In the SM,  all masses are attributed to a single parameter of the vacuum expectation value (VEV),  $\langle H \rangle$,  of 
the hypothetical elementary particle, the Higgs boson,which triggers the spontaneous symmetry breaking (SSB). The VEV  simply picks up the mass scale of the input parameter $M_0$ 
which is tuned to be tachyonic ($M_0^2<0$) in such a way that  
$\langle H \rangle \simeq 246 \,{\rm GeV}$  (``naturalness problem''). 
As such SM does not explain the Origin of Mass. 
Particle theorists looking desperately beyond the SM have been fighting  on this central problem over 30 years without decisive experimental
information.  Now we are facing a new era that LHC experiments  will tell us which
theory is right. 

It should be recalled that the very concept of SSB was created by the 2008 Nobel prize work of Nambu~\cite{Nambu, Nambu:1961tp} in a concrete form of  the dynamical symmetry breaking (DSB) where the nucleon mass was dynamically generated via Cooper pairing of (then elementary) nucleon and anti-nucleon, ``nucleon condensate'',  based on the Bardeen-Cooper-Schrieffer (BCS) analogue of superconductor: Accordingly, there appeared pions as massless Nambu-Goldstone (NG) bosons which were dynamically generated to be nucleon composites in the same sense as in the Fermi-Yang~\cite{Fermi:1959sa}/Sakata model~\cite{Sakata:1956hs}. 
{\it Thus the SSB was born as DSB!} Before advent of the concept of SSB, low energy hadron physics was well described by 
the effective theory of  Gell-Mann-Levy (GL) linear sigma model~\cite{GellMann:1960np}with an elusive scalar boson, the sigma meson, 
which was  simply assumed to have negative mass squared. 
Actually,  the GL linear sigma model Lagrangian is  a model formally equivalent to the SM Higgs Lagrangian, with the Higgs boson being the counterpart of the sigma meson .
The real physical meaning of this mysterious tachyonic mode was thus revealed
 as the BCS instability where attractive forces (effective four-fermion interactions) between nucleon and anti-nucleon give rise to the
nucleon Cooper paring (tachyonic bound state) which changes the vacuum from the original (free) one into  the true one
having no manifest symmetry.

The Nambu's theory for the origin of mass of nucleon (then the ``elementary particle'') was later developed into DSB in the underlying microscopic theory, QCD,  where the gluonic attractive forces again generate the Cooper paring of quark and antiquark (instead of nucleon and anti-nucleon), the quark condensate $\langle \bar q q \rangle$, which then gives rise to  the BCS instability and the dynamical mass of quarks: Pions are now composites of quarks instead of nucleons. Hence Nambu's idea was established in a deeper level of matter. 
Note that the nucleon mass in the Nambu's theory  is originated from the explicit  mass scale carried by the dimensionful coupling in the four-fermion theory, while in QCD (with massless quarks) there is no mass scale at classical level: The intrinsic mass scale in QCD, $\Lambda_{\rm QCD}$, {\it arises from the scale anomaly quantum mechanically}, which  manifests itself  in the {\it running of the gauge coupling}.  This is a salient feature of the mass generation of the gauge theory.

 Technicolor (TC)~\cite{TC} is an attractive idea to account for the Origin of Mass 
 without introducing ad hoc Higgs boson and tachyonic mass parameter:  The mass arises {\it dynamically} from the condensate of the techni-fermion and the anti-techni-fermion pair $\langle \bar F F \rangle$ which is triggered by the attractive gauge forces between the
 pair analogously to the quark-antiquark condensate $\langle \bar q q \rangle $
  in QCD.    
  The dynamically generated mass scale for the $W/Z$ boson mass is characterized by  the dynamical mass of the techni-fermion,  $m_{_{F}}={\cal O} ({\rm TeV})$ which is a universal scale of techni-hadron mass $M_{\rm TH} ={\cal O}(m_{_{F}})$  and the techni-condensate: $\langle \bar F F \rangle \sim - m_{_{F}}^3$. Actually,  $m_{_{F}}$
 picks up the intrinsic mass scale  $\Lambda_{\rm TC}$ of the theory 
 (analogue of $\Lambda_{\rm QCD}$ in QCD) already generated by the {\it scale anomaly through quantum effects  (``dimensional transmutation'')}  in the gauge theory which is {\it scale-invariant at classical level} (for massless flavors): 
 \beq
 m_{_{F}} \sim \Lambda_{\rm TC} = \mu \cdot \exp \left( - \int^{\alpha(\mu)} \frac{d\alpha}{\beta(\alpha)} \right) \,, 
\label{intrinsic}
\eeq 
where $\Lambda_{\rm TC}$ is independent of the renormalization point $\mu$, $\frac{d \Lambda_{\rm TC}}{d \mu}=0$,  and the running (scale-dependence) of the coupling constant $\alpha(\mu)$, with non-vanishing beta function $ \beta(\alpha) \equiv \mu \frac{d \alpha(\mu)}{d \mu} \ne 0$,
 is a manifestation of the scale anomaly. 
Thus the {\it Origin of Mass is eventually the quantum effect (scale anomaly)} in this picture, with the 
 scale symmetry broken explicitly at the scale of ${\cal O} (\Lambda_{\rm TC})$ without remnant of the scale symmetry at all.
Note that the intrinsic scale $\Lambda_{\rm TC}$
 can largely be separated from  the fundamental scale, the Planck scale $\Lambda_{Pl}$ 
 through logarithmic running  (``naturalness''):
 \beq
{\rm Naturalness}({\rm QCD/ScaleUp}): \quad  M_{\rm TH}= {\cal O} (m_{_{F}}) = {\cal O} (\Lambda_{\rm TC}) ={\cal O} ( {\rm TeV}) \ll \Lambda_{Pl} \,.
 \label{scaleup}
  \eeq

However, the original version of TC~\cite{TC}, a naive scale-up version of QCD,  was dead due to the excessive 
flavor-changing neutral currents (FCNC).   In order to give mass to the quarks/leptons not just to the $W/Z$ boson,
we need to introduce another scale, say $\Lambda_{\rm ETC}$, typically through the extended TC (ETC) model~\cite{Dimopoulos:1979es}   
\footnote{
Such a scale can be introduced by other models, for instance a composite quark/lepton/techni-fermion model~\cite{Yamawaki:1982tg}
.
} :  $m_{\rm q/l}\sim -\langle \bar F F\rangle/\Lambda_{\rm ETC}^2\sim m_{_{F}}^3/\Lambda_{\rm ETC}^2$. This also induces FCNC roughly of order $1/\Lambda_{\rm ETC}^2$, which is constrained by the experiments as $\Lambda_{\rm ETC} > 10^3-10^4$ TeV and hence reproduces only $10^{-3}$ times the realistic value for   
 $m_{\rm q/l}$.

 It  was resolved long time ago by the Walking TC (WTC)~\cite{Yamawaki:1985zg}, initially dubbed ``Scale-Invariant Technicolor'', based  on the SSB solution of the ladder Schwinger-Dyson (SD) equation with 
 {\it non-running} ({\it scale invariant/conformal}) gauge coupling, $\alpha(p) \equiv \alpha$, which we found
 gives rise to  the large anomalous dimension, 
 \beq
 \gamma_m=1\,,
 \eeq
 when the dynamical mass generation $m_{_{F}}\ne 0$ takes place for strong coupling $\alpha>\alpha_{\rm cr} (={\cal O}(1))$, thus enhancing the techni-fermion condensate 
 $\langle \bar F F\rangle|_{\mu}=Z_m^{-1} \langle \bar F F\rangle|_{\mu=m_{_{F}}}$,
 $\langle \bar F F\rangle|_{\mu=m_{_{F}}} \sim -m_{_{F}}^3$, by the factor $Z_m^{-1} =Z_m^{-1} (\mu/m_{_{F}})=(\mu/m_{_{F}})^{\gamma_m}=\mu/m_{_{F}}$ such that $\Lambda_{\rm ETC}/m_{_{F}}\sim 10^3$ for $\mu =\Lambda_{\rm ETC}$.
   (A solution to the FCNC problem by the large anomalous dimension was suggested earlier by simply assuming the existence of 
 a large anomalous dimension without any concrete dynamics and concrete  
 value of the anomalous dimension~\cite{Holdom:1981rm}). It was noted~\cite{Yamawaki:1985zg,Bando:1986bg} that the coupling ($> 
 \alpha_{\rm cr}$) actually does become 
   {\it running slowly (``walking'') nonperturbatively}  a la Miransky~\cite{Miransky:1984ef}: $\alpha=\alpha(Q)$ 
for $m_{_{F}} <Q<\Lambda_{\rm ETC}\,\, (Q^2\equiv -p^2>0) $, with {\it nonperturbative} beta function,~\cite{Bardeen:1985sm}
$\beta^{^{\rm NP}}(\alpha) =
-(2 \alpha_{\rm cr}/\pi)
\cdot \left(\alpha/
\alpha_{\rm cr}
-1\right)^{3/2}
$,   
yielding a {\it non-perturbative scale anomaly} at the scale $m_F$  dynamically generated by the SSB.
Subsequently, a similar FCNC solution   
 was  discussed without 
notion of anomalous dimension and scale invariance.~\cite{Akiba:1985rr}.   The {\it WTC also predicted a Techni-dilaton (TD)}~\cite{Yamawaki:1985zg,Bando:1986bg}, a pseudo Nambu-Goldstone (NG) boson of the approximate scale symmetry, which is a composite Higgs, a scalar $\bar F F$ bound state,
 behaving similarly to the SM Higgs and will be most relevant to the LHC physics as we will discuss in this talk. 
  (For reviews of WTC see
Ref. \cite{Hill:2002ap}).
  
 The mass generation due to such a scale-invariant (conformal) dynamics 
 takes the form of essential-singularity scaling, Miransky scaling~\cite{Miransky:1984ef}, 
\begin{equation}
m_{_{F}} \sim 
\Lambda \cdot \exp 
\left(
 - \int^{\alpha(\Lambda)} \frac{d\alpha}{\beta^{^{\rm NP}}(\alpha) }
\right) 
\sim
\Lambda \cdot \exp 
\left(-
\frac{
\pi
}{
\sqrt{
\frac{
\alpha}{
\alpha_{\rm cr}
}
-1}
}
\right)
  \ll \Lambda \,,
\label{Mscaling}
\end{equation}
in a way to ensure a large natural hierarchy $m_{_{F}} \ll \Lambda$ (for $\alpha \simeq \alpha_{\rm cr}$). This is  characterized by the ``{\it conformal phase transition}''~\cite{Miransky:1996pd}. 
Thus the essence of the WTC is a model setting of walking (scale-invariant) coupling $\alpha(p)\simeq const. \simeq \alpha_{\rm cr} $ as an input (perturbative) coupling in the SD equation, 
which results in non-perturbatively walking coupling for the wide energy region $m_{_{F}}  < Q <\Lambda
\,\,(m_{_{F}}\ll \Lambda)$ for $\Lambda= \Lambda_{\rm ETC}$.

Such a situation is actually realized~\cite{Lane:1991qh,Appelquist:1996dq}~\cite{Miransky:1996pd} by the two-loop perturbation 
 in the large 
$N_f$ QCD which has the Caswell-Banks-Zaks (CBZ) IR fixed point ~\cite{Caswell:1974gg} $\alpha_*$ in the beta function:  $\beta^{(2-loop)}(\alpha) =- b \alpha^2 (1-\alpha/\alpha_*) $, ($b>0$). 
Due to the CBZ IR fixed point,  the in-put  coupling in the SD equation is {\it almost non-running in the IR region: $\alpha (Q)\simeq \alpha_*$ for  $0<Q< \Lambda_{\rm TC}$},  
while
 it runs asymptotically free and diminishes rapidly 
in the same way as the ordinary QCD in the UV region:  $ \alpha (Q)\sim  1/\ln(Q/\Lambda_{\rm TC})$ for $Q>\Lambda_{\rm TC}$, where
$\Lambda_{\rm TC}$ is the intrinsic scale, a two-loop analogue of  $\Lambda_{\rm QCD}$,  defined by Eq.(\ref{intrinsic}) with the two-loop beta function
$\beta(\alpha) \Rightarrow \beta^{(2-loop)}(\alpha)$. 
Thus  the situation is  similar to the original model~\cite{Yamawaki:1985zg},  with {\it  $\Lambda_{\rm TC}$ playing a role of the UV cutoff $\Lambda=\Lambda_{\rm ETC}$: $\Lambda_{\rm TC}\sim \Lambda_{\rm ETC}$}.~\footnote{
For $Q>\Lambda_{\rm TC}\sim \Lambda_{\rm ETC}$ the  WTC model  no longer makes sense as 
the same theory but becomes a part of a larger model like ETC, though. Dynamical origin of the scale $\Lambda_{\rm ETC}$ is a 
separate issue which may be the scale of tumbling or further compositeness arising eventually from the scale anomaly in the more
fundamental gauge theory.  
}

Salient feature of the WTC is the large hierarchy of techni-hadron masses and $\Lambda_{\rm TC}$:  $M_{\rm TH} ={\cal O} (m_{_{F}}) \ll \Lambda_{\rm TC}$,
  which is naturally realized as in Eq. (\ref{Mscaling}) by the scale-invariant dynamics.  This is contrasted to the QCD-scale up TC where there is no hierarchy between the techni-hadron mass 
  and the intrinsic scale $\Lambda_{\rm TC}$ which is a typical IR scale instead of UV scale: $M_{\rm TH} ={\cal O}(m_{_{F}})  ={\cal O}(\Lambda_{\rm TC})$.

Even more striking feature of the WTC  is the  {\it ``Techni-dilaton (TD)''}~\cite{Yamawaki:1985zg,Bando:1986bg},
a naturally light scalar $\bar F F$ composite  Higgs as mentioned above.
As a pseudo NG boson, the TD mass $M_{\rm TD}$ is even smaller than those of  all other  techni-hadrons: $M_{\rm TD} < M_{\rm TH}$. This is against a long-standing folklore that the TC is a ``Higgsless'' model having no light scalar composite.
Such a folklore is applied only to the original TC as a simple scale-up of the QCD 
where the  coupling is {\it running already at perturbative level},  with the scale symmetry badly broken 
for all energy region and there is no remnant of the scale symmetry. Thus,  besides that the  hierarchy $\Lambda_{\rm TC} 
 \ll \Lambda_{Pl}$ as in QCD
scale-up 
in Eq.(\ref{scaleup}), we have additional natural hierarchies related with the scale symmetry:  
\begin{eqnarray}
 {\rm Naturalness}\,\, ({\rm WTC}) : \quad M_{\rm TD} < M_{\rm TH} 
                                                                     ={\cal O} ( m_{_{F}})
                                                                                                              \ll  \Lambda_{\rm TC} \ll \Lambda_{Pl}\,,
\end{eqnarray}
with a model setting $\Lambda_{\rm TC} \sim \Lambda_{\rm ETC}$. 
From various calculations~\cite{Shuto:1989te,
    Harada:2003dc}
 related with the ladder approximation we suggested~\cite{Yamawaki:2010ms}
 that the TD mass in a typical TC model (one-family model)  is (up to large uncertainty in the ladder-like calculations):
\beq
M_{\rm TD} \simeq 500 - 600 \,\, {\rm GeV}\,.
\eeq

In this talk we argue~\cite{Matsuzaki:2011ie} that such a TD as a composite Higgs
can be soon discovered by the LHC experiment~
\cite{ATLAS}.~\footnote{
After the Symposium, LHC announced some excess around $125 {\rm GeV}$~\cite{ATLAS1213}, which happen to be consistent with 
the techni-dilaton!~\cite{Matsuzaki:2012gd}  Such a lower mass of the techni-dilaton is in fact protected by the scale symmetry:
The quadratic divergence loop corrections are $\delta M_{\rm TD}^2 \sim \mu^2/(4\pi)^2 < m_{_{F}}^2/(4\pi)^2 \ll m_{_{F}}^2$, with those from $m_{_{F}} < \mu<\Lambda_{\rm TC}$ being  highly suppressed by the scale symmetry.}
We will also describe the related composite Higgs boson in various models of dynamical symmetry breaking with large anomalous dimension (For reviews see Ref.~\cite{Yamawaki:1996vr}), namely a class of composite Higgs models based on the walking gauge dynamics having large anomalous dimension characteristic to the conformal UV/IR fixed point:
Strong-ETC TC 
with  $1<\gamma_m<2$~\cite{Miransky:1988gk},  
Top Quark Condensate Model~\cite{Miransky:1988xi} with  $\gamma_m \simeq 2$, and their variants. 
 They will be tested soon in the on-going  LHC experiments.

\section{Walking Technicolor}
\label{aba:sec2}
The WTC
 is the model with dynamical mass generation $m_F$ with
large anomalous dimension:~\cite{Yamawaki:1985zg} 
\beq
\gamma_m \simeq 1 \,.
\label{gammam1}
\eeq
\begin{wrapfigure}{r}{6.6cm}
\vspace{-0.5cm}
\includegraphics[width=6cm]{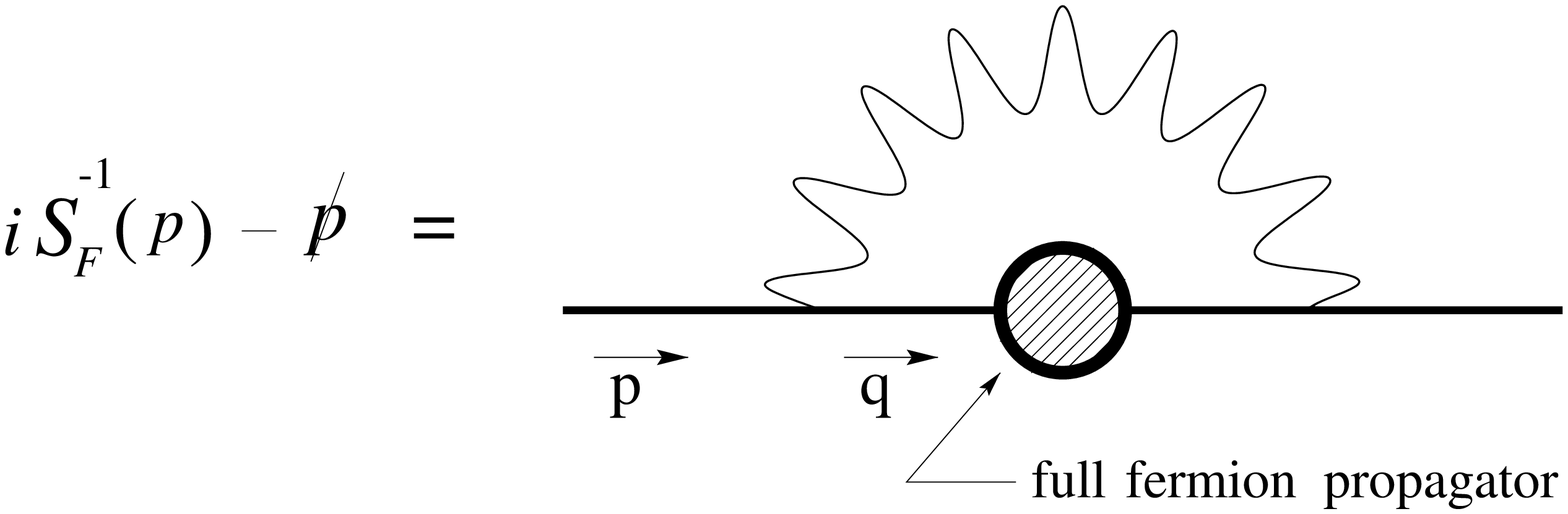}
\caption{Graphical expression of the SD equation 
        in the ladder approximation.}
\label{fig:SDeq}
\end{wrapfigure}
The model was proposed  based on  the SSB solution of the ladder Schwinger-Dyson (SD) equation (Fig. \ref{fig:SDeq})  for the fermion full propagator $S_F(p)$ parameterized as
$i S_{F}^{-1}(p) = A(p^2) \fsl{p} - B(p^2)$, 
with non-running
(scale-invariant) gauge coupling 
$\alpha( Q) \equiv \alpha > \alpha_c$ ($Q^2\equiv -p^2>0$). 
It was shown~\cite{Maskawa:1974vs}  that the SSB solution exists only for strong coupling $\alpha>\alpha_{\rm cr} ={\cal O} (1)$. 

The asymptotic  form of the SSB solution of the fermion mass function
 $\Sigma(Q) =B(p^2)/A(p^2)$ ($\Sigma(m_{_{F}})=m_{_{F}}$)
 in Landau gauge ($A(p^2)\equiv 1$) reads~\cite{Maskawa:1974vs,Fukuda:1976zb}, 
 \begin{eqnarray}
\Sigma (Q) \sim 1/Q \quad (m_{_{F}} < Q <\Lambda)
 \,, 
\label{asymp}
\end{eqnarray}
which we found~\cite{Yamawaki:1985zg}  implies a large value of the anomalous dimension
$\gamma_m  
= 1,  
$
to be compared  with the operator product expansion (OPE), $\Sigma (Q) \sim 1/Q^2\cdot (Q/m_{_{F}})^{\gamma_m}$.
Accordingly, we had a linearly divergent  condensate   $\langle \bar{F} F \rangle_{\Lambda}=   Z_m^{-1}\cdot \langle \bar{F} F \rangle_{m_{_{F}}} \sim -
(\Lambda/m_F)\cdot  
m_{_{F}}^3$, with the (inverse) mass renormalization constant  being $Z_m^{-1} = (\Lambda
/m_{_{F}}
)^{\gamma_m} =
\Lambda
/m_{_{F}}
> 10^3$ for $ \Lambda=\Lambda_{\rm ETC}>10^3 m_{_{F}}$, which in fact  yields the desired enhancement:  
We actually obtained  $
 m_{q/l} \sim m_{_{F}}^2/\Lambda_{\rm ETC}
$\cite{Yamawaki:1985zg}, 
which is compared with the  QCD scale-up TC $m_{\rm q/l} \sim m_{_{F}}^3/\Lambda_{\rm ETC}^2$.

The model~\cite{Yamawaki:1985zg} was actually formulated in  terms of the Miransky's nonperturbative renormalization~\cite{Miransky:1984ef} of the
SSB solution  
which  takes the  {\it essential singularity} form of Eq.(\ref{Mscaling}) (Miransky scaling):~\cite{Fukuda:1976zb, Miransky:1984ef}   
 \beq
 m_{_{F}} \sim 
 \Lambda \, \exp \left(- \pi/\sqrt{\alpha/\alpha_{\rm cr} -1}
\right) \, .
\label{Miransky}
\eeq  
where the critical value $\alpha_{\rm cr}$ reads:~\cite{Fukuda:1976zb}
 \beq
 C_2(F) \cdot\alpha_{\rm cr}= \frac{\pi}{3} 
 \label{critical}
 \eeq
in the $SU(N_{\rm TC}) $ gauge theory,  where $C_2(F) = (N_{\rm TC}^2-1)/2N_{\rm TC}
$ is the quadratic Casimir of
the techni-fermion representation. 
Once $m_{_{F}}$ 
is dynamically generated in such a way,  the coupling does depend on $\Lambda/m_{_{F}}$, and no longer remains constant but does start {\it walking}  with the {\it nonperturbative} beta function:~\cite{Bardeen:1985sm}
 \beq
\beta^{^{\rm NP}}(\alpha)= 
\frac{\partial \alpha
}{\partial \ln (\Lambda/m_{_{F}})}= 
-
\frac{
2\pi^2 \alpha_{\rm cr} 
}
{
\ln^3 
(
\Lambda/
m_{_{F}}
)
}
=
-\frac{2\alpha_{\rm cr}}{\pi}\left( \frac{\alpha}{\alpha_{\rm cr}}-1 \right)^{3/2}\, ,
\label{Miranskybeta}
\eeq 
where
 the critical coupling $\alpha_{\rm cr}$ was identified with a nontrivial UV fixed point 
 $\alpha=\alpha(\Lambda/m_{_{F}})  \rightarrow \alpha_{\rm cr}$
as $\Lambda/m_{_{F}} \rightarrow \infty$. 
This reflects {\it  explicit breaking} of the scale symmetry (nonperturbative scale anomaly) due to the generated mass scale $m_F$ which is the very origin of
{\it spontaneous breaking} of the scale symmetry. 
Although the IR scale $m_{_{F}}$ is originated from the UV scale (fundamental/intrinsic scale) $\Lambda$ of the theory as in Eq.(\ref{Miransky}), this {\it scale anomaly characterized by the scale $m_{_{F}}$ is persistent}, even if we removed the
UV scale $\Lambda \rightarrow \infty$ (``perturbatively complete scale-invariant limit''). 
Note that  
the {\it essential singularity} corresponds to  
{\it multiple zero} of $\beta^{^{\rm NP}}(\alpha)$ as seen from Eq.(\ref{Mscaling}), which is {\it never realized in the perturbative calculations}.~\footnote{
Linear zero of the perturbative beta function, $\beta(\alpha) \sim (\alpha-\alpha_{\rm cr})^1$,  never reproduces the essential singularity scaling, as is evident from 
Eq.(\ref{intrinsic}).  
}

The essential feature of the above is precisely what happens in the 
modern version ~\cite{Lane:1991qh,Appelquist:1996dq,Miransky:1996pd}  of the WTC
 based on the CBZ IR fixed point~\cite{Caswell:1974gg}  of  the large $N_f$ QCD,   the QCD-like theory with many  flavors $N_f  \, (\gg N_{\rm TC})$ of  massless 
 techni-fermions.~\footnote{
For WTC based on higher representation/other gauge groups see, e.g.,  Ref. \cite{Sannino:2004qp} 
}  .
The  two-loop beta function is given by
 $\beta^{(2-loop)}(\alpha)  
  = -b \alpha^2(\mu) - c \alpha^3(\mu) =- b \alpha^2 (1-\alpha/\alpha_*);
$
$  b = \left( 11 N_{\rm TC} - 2 N_f \right)/(6 \pi)$, 
 $ c =\left[ 34 N_{\rm TC}^2 - 10 N_f  N_{\rm TC} 
      - 3 N_f  (N_{\rm TC}^2 - 1)/N_{\rm TC} \right] /(24 \pi^2)$ .
When $b>0$ and $c<0$, i.e.,  $  N_f^* < N_f < \frac{11}{2} N_{\rm TC} $ 
($N_f^\ast \simeq 8.05$ for $N_{\rm TC} = 3$), there exists an IR fixed point (CBZ  IR fixed point)
  at $\alpha=\alpha_*$, $\beta(\alpha_*)=0$, where
\begin{equation}
  \alpha_\ast =\alpha_*(N_{\rm TC}, N_f)  = - b/c .
\label{eq:alpha_IR}
\end{equation}
The intrinsic scale $\Lambda_{\rm TC}$, a two-loop analogue of the $\Lambda_{\rm QCD}$, may be chosen as $\alpha(\Lambda_{\rm TC})=1/(1+1/e) \alpha_* \simeq 0.7 \alpha_*$, so that 
the two-loop coupling is almost non-running in the IR region 
: $\alpha(Q) \simeq \alpha_*$ ($
Q\ll\Lambda_{\rm TC}$), while it runs as in the ordinary QCD in the UV region; $\alpha(Q) \sim 1/\ln (Q/\Lambda_{\rm TC})$  ($Q\gg \Lambda_{\rm TC}$).    Thus $\Lambda_{\rm TC}$ plays a role of cutoff $\Lambda \,
(\sim \Lambda_{\rm ETC})$  in the original model~\cite{Yamawaki:1985zg} .  Note that $\alpha_* =\alpha_*(N_f,N_{\rm TC} ) \rightarrow 0$ as $N_f \rightarrow  11 N_{\rm TC}/2$  $ (b \to 0)$ and hence there exists a certain range $N_f^{\rm cr}<N_f < 11 N_{\rm TC}/2$  (``Conformal Window'')  satisfying $\alpha_* < \alpha_{\rm cr}$, where the gauge coupling $\alpha (p) \, (< \alpha_* < \alpha_{\rm cr})$ is not
strong enough to trigger the SSB. The $N_f^{\rm cr}$ such that  $\alpha_*(N_{\rm TC}, N_f^{\rm cr}) = \alpha_{\rm cr}$ may be evaluated by using  the value of
$\alpha_{\rm cr}$
 from  the ladder SD equation Eq.(\ref{critical}):~\cite{Appelquist:1996dq}  
$N_f^{\rm cr} \simeq 4 N_{\rm TC}$ ($ =12$ for $N_{\rm TC}=3$)
\footnote{
The value should not be taken seriously, since $\alpha_\ast=\alpha_{\rm cr}$ is of  $\cal {O} $(1) and
the perturbative estimate of  $\alpha_*$  is not so reliable.
Lattice simulations still  suggest diverse results as to $N_f^{\rm cr}$. 
}.

Here we are interested in the SSB phase slightly outside of the conformal window, 
$0< \alpha_\ast - \alpha_{\rm cr}\ll 1$ ($N_f \simeq N_f^{\rm cr}$).  
We may use the same equation as the ladder SD  equation  with $\alpha(Q) \simeq {\rm const.} =\alpha_*$, yielding 
 the
same  
result, Eqs.(\ref{gammam1},\ref{asymp},\ref{Miransky}):
\begin{eqnarray}
\Sigma(Q) \sim 1/Q \,,\quad 
\gamma_m \simeq  1,  \quad (m_{_{F}} < Q <\Lambda_{\rm TC} (\sim \Lambda_{\rm ETC}))\,.
\end{eqnarray}
\begin{equation}
m _{_{F}}\sim 
\Lambda_{\rm TC} \, \exp \left(- \pi/\sqrt{\alpha_*/\alpha_{\rm cr} -1}
\right) \ll  \Lambda_{\rm TC}   \quad ( \alpha_\ast \simeq \alpha_{\rm cr})\, ,
\label{Appel}
\end{equation}where the cutoff $\Lambda\, (=\Lambda_{\rm ETC})$ was identified with $\Lambda_{\rm TC} $.  

\begin{wrapfigure}{r}{5.0cm}
\vspace{-0.5cm}
\includegraphics[width=5cm]{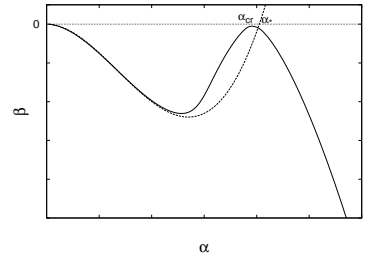}
\caption{
Shape of the full $\beta$ function:  $\sim \beta^{^{\rm NP}}(\alpha)$ ($\alpha>\alpha_{\rm cr}$),
$\beta^{(2-loop)}(\alpha)$  ($\alpha<\alpha_{\rm cr}$). 
  The bold solid and dashed curves correspond to the 
  full $\beta$ function and the two-loop one, $\beta^{(2-loop)}(\alpha)$,  respectively.}
\end{wrapfigure}
Eq.(\ref{Appel}) implies 
a nonperturbative beta function $\beta^{^{NP}} (\alpha)$  of the form Eq.(\ref{Miranskybeta}) for $\alpha=\alpha_*=\alpha (\Lambda_{\rm TC}/m_{_{F}})$,  with  $\alpha_{\rm cr}$ regarded as  a UV fixed point  in the  limit $\Lambda_{\rm TC}/m_{_{F}} \rightarrow \infty$.
Such a beta function has a {\it multiple zero} corresponding to the essential singularity as noted before. 
Thus the two-loop beta function $\beta^{(2-loop)}(\alpha)$  is to be replaced by the nonperturbative one, $\beta (\alpha) \sim \beta^{^{NP}} (\alpha)$, as in the original model\cite{Yamawaki:1985zg},  whereas it is essentially operative, $\beta (\alpha) \sim \beta^{(2-loop)}(\alpha)$,  in the UV region $Q> \Lambda_{\rm TC}$ ($\alpha<\alpha_{\rm cr}$).
 
 Hence the full beta function  
  within the SD equation analysis may be depicted as in Fig.~\ref{betafull}.~\cite{Hashimoto:2010nw} \label{betafull}
  This should be tested by the fully non-perturbative studies like lattice simulations.
A possible phase diagram (Fig 3 of Ref.\cite{Miransky:1996pd}) of  the  large $N_f$ QCD on the lattice  is  also waiting for the test by simulations.

\section{S Parameter Constraint}

Now we come to the next problem of TC, so-called $S,T,U$ parameters \cite{PeskinTakeuchi} measuring possible new physics in terms of  the deviation of the LEP precision experiments  from the SM.  
The most straightforward computation of the $S$ parameter for the large $N_f$ QCD
is that based on the SD equation and (inhomogeneous) 
Bethe-Salpeter (BS) equation in the ladder approximation. \cite{Harada:2005ru}
The results show that  there is a tendency of the $S$ getting reduced when
approaching the conformal window $\alpha_\ast/\alpha_{\rm cr}\searrow 1$ 
($N_f \nearrow N_f^{\rm cr}$), and at the present technical computational limit  $\alpha_*/\alpha_{\rm cr}\sim 1.13$
 it reads $S \sim 0.056 (N_{\rm TC} N_D)$ for $N_D$ doublets, about a half of that in the ordinary QCD,
 $S_{\rm QCD} \sim 0.11 N_c\,\, (N_D=1)$. The result is 
barely consistent with the experiments  for $N_{\rm TC} =2,  N_D=1$.   
 Highly desirable is 
 the computation closer to $\alpha_*/\alpha_{\rm cr}=1$.

As another approach 
the reduction of $S$ parameter in the WTC has  been argued~\cite{Hong:2006si,Haba:2008nz} in  a version of  the hard-wall type bottom up holographic 
  QCD~\cite{Erlich:2005qh}
deformed to WTC  by  tuning a parameter
to simulate the large anomalous dimension $\gamma_m \simeq 1$. 
We find~\cite{Haba:2008nz}    $S/N_D\simeq 8\pi (F_\pi/M_\rho)^2 =4\pi/g^2$ for $N_{\rm TC}=3$,
where $M_{\rho}^2=a F_\pi^2 g^2$ 
($a \simeq 2$) and $g$  is the gauge coupling of the hidden local symmetry~\cite{Bando:1984ej}.   
Although the results contain full contributions from an  infinite tower of the vector/axial-vector Kaluza-Klein modes (gauge
bosons of hidden local symmetries) of the 5-dimensional gauge bosons,
the resultant value turned out  close to that of a single $\rho$ meson dominance.  
This implies that as far as the pure TC dynamics (without ETC dynamics, etc.) is concerned,
the $S$ parameter can be reduced only by tuning $F_\pi/M_\rho$ very small, namely techni-$\rho$ mass very large to several TeV region.  
Curiously enough,  
when we calculate  $F_\pi/M_\rho$ from the SD/homogeneous BS equations~\cite{Harada:2003dc} and 
$S$ from the SD/inhomogeneous BS equation~\cite{Harada:2005ru} both   
in the ladder approximation,
a set of the calculated values $(F_\pi/M_\rho, S/N_D)\simeq (0.08,0.17)$  for $N_{\rm TC}=3$ lies on the line of the holographic result~\cite{Haba:2008nz}.

\section{Techni-dilaton}
 Now we come to the discussions of Techni-dilaton (TD).
Existence of approximate scale invariance for for wide region $m_F < Q < \Lambda_{\rm TC}$ 
between two largely separated scales, 
$m_{_{F}}\ll \Lambda_{\rm TC}$, 
is the most important feature of WTC near conformality,  in sharp contrast to 
the ordinary QCD 
(in the chiral limit)  
where there is no scale invariant region and all the mass parameters $M$  
are of
order of a single scale parameter of the theory $\Lambda_{\rm QCD}$, $M=  {\cal O} (\Lambda_{\rm QCD})$. 
The intrinsic scale $\Lambda_{\rm TC}$ is related with the scale anomaly corresponding to
the {\it perturbative running} effects
of the coupling $\alpha(Q)$ for $Q>\Lambda_{\rm TC}$, with the ordinary  two-loop beta function 
$\beta^{(2-loop)}$
in the same sense as in QCD. 
$\VEV{\partial^\mu D_\mu}^{^{({\rm pert.})}}= \VEV{\theta^{\mu}_{\mu}}^{^{({\rm pert.})}} 
= \frac{\beta(\alpha)}{4 \alpha^2} \GB \Big|^{^{({\rm pert.})}}=- {\cal O} (\Lambda_{\rm TC}^4). $ 

In the WTC,  on the other hand,  there exists another scale $m_{_{F}} \, (\ll \Lambda_{\rm TC})$ in addition to $\Lambda_{\rm TC}$, where the largely separate scale $m_{_{F}}$  is related with a totally different scale anomaly associated with the
{\it nonperturbative running} of the coupling due to the dynamical generation of $m_F$ as in Eq.(\ref{Miranskybeta}), 
which does exist 
even in the idealized case with perturbatively scale-invariant  coupling 
$\alpha^{^{pert.}}(Q) \equiv \alpha ( >\alpha_{\rm cr})$  
for entire region $0<Q< \Lambda_{\rm TC} \rightarrow \infty$. Note that there is {\it no idealized  limit where the TD  becomes exactly massless} to be a true NG boson,  in sharp contrast to the chiral symmetry breaking. The SSB of the scale symmetry as well as the chiral symmetry  is triggered by the dynamical generation of $m_F$,
which however is the very cause of the nonperturbative running of the coupling, namely the nonpertubative scale anomaly: The scale symmetry is always {\it broken explicitly as well as spontaneously}.~\cite{Haba:2010hu,Hashimoto:2010nw}

The {\it non-perturbative} scale anomaly reads~\cite{Miransky:1996pd} 
\beqs
\VEV{\partial^\mu D_\mu}^{^{\rm NP}} 
= \VEV{\theta^{\mu}_{\mu}}^{^{\rm NP}} 
= \frac{\beta^{^{\rm NP}} (\alpha)}{4 \alpha^2} \GB^{^{\rm NP}}  =-   m_{_{F}}^4\cdot \kappa_V\, \frac{N_f N_{\rm TC}}{2\pi^2}\,,
\label{npanomaly}
\eeqs
where $\langle \cdots\rangle^{^{\rm NP}}$ is the quantity without perturbative contributions   
$\langle \cdots\rangle^{^{\rm NP}} \equiv  \langle \cdots\rangle - \langle \cdots\rangle^{^{({\rm pert.})}}$. Here $\kappa_V=0.76$~\cite{Hashimoto:2010nw}
 for  the two-loop coupling ($N_f \simeq N_f^{\rm cr}$), while  
$\kappa_V=8/\pi^2\simeq 0.81$~\cite{Miransky:1989qc}   for the non-running coupling in the SD equation.
All the techni-fermion bound states have mass $M_{\rm TH} ={\cal O} (m_{_{F}})$
~\cite{Chivukula:1996kg}, while there are no light bound states
in the symmetric phase (conformal window) $\alpha_*<\alpha_{\rm cr}$, a characteristic feature of the conformal phase transition~\cite{Miransky:1996pd}.
The TD is associated with the latter scale anomaly and should also have a mass $M_{\rm TD}={\cal O}(m_F) (\ll \Lambda_{\rm TC})$.
It actually turns out that $M_{\rm TD}$ is  even smaller than other techni-hadron mass:
$M_{\rm TD} <M_{\rm TH} $.

To be concrete we estimate $m_F$ related to $v=246 \, {\rm GeV}$ as
\begin{equation} 
  F_\pi^2 =N_D v^2 = \kappa_F^2 \frac{N_{\rm TC}}{4 \pi^2} m_F^2 
\,,   
\label{PS}
\end{equation}
where $\kappa_F\simeq 1.5$ from the Pagels-Stokar formula for the mass function at 
ladder-like criticality $N_f \simeq N_f^{\rm cr}\simeq 4N_{\rm TC}$ and
$N_D$ is the number of weak doublets ($N_D=4$ for one-family model and $N_D=1$ for one-doublet model)~\cite{Hashimoto:2010nw}.
We take $N_f=2N_D + N_{\rm EW-singlet}$, where $N_{\rm EW-singlet}$ is the number of electroweak singlet
techni-fermions which only contribute to the walking  coupling. Thus we have rough estimate
\beq
m_{_{F}} \simeq 1 {\rm TeV} /\sqrt{N_D N_{\rm TC}}\,.
\label{mF}
\eeq

\subsection{Calculation from Gauged NJL model in the ladder SD equation~\cite{Shuto:1989te}}
More concretely, the mass of TD was estimated~\cite{Yamawaki:2010ms} by reinterpretation of the old results on the scalar bound state mass by various methods:   
The first method~\cite{Shuto:1989te}  was based on 
the  ladder SD equation for the gauged NJL model which well simulates~\cite{Miransky:1996pd} the 
conformal phase transition in the large $N_f$ QCD.  We find:
\beqs
M_{\rm TD} \simeq 
\sqrt{2} m_{_{F}} \, .
\label{PCDCmass}
\eeqs

\subsection{Straightforward Calculation from Ladder SD and BS equations~\cite{Harada:2003dc}}

Also a straightforward calculation of the mass of TD, the scalar bound state, 
was made  in the
vicinity of the CBZ-IR fixed point  in the large $N_f$ QCD, based on the coupled use of the ladder SD equation and ({\it homogeneous}) BS equation.
All the bound states masses $M_{\rm TH}$ as well as  $F_\pi$ are of order ${\cal O}(m_{_{F}})$ and $M_{\rm TH}/\Lambda_{\rm TC}, F_\pi/\Lambda_{\rm TC}\to 0$, when approaching 
the conformal window  $\alpha_*\to \alpha_{\rm cr} \,\, (N_f \to N_f^{\rm cr})$ such that $m_{_{F}}/\Lambda_{\rm TC} \to 0$.  
Near the conformal window ($N_f \nearrow N_f^{\rm cr}$)  the calculated values are  $M_\rho/F_\pi \simeq 11, M_{a_1}/F_\pi  \simeq 12$ (near degenerate !).
On the other hand, the scalar mass sharply drops near 
the conformal window,   $M_{\rm TD}/F_\pi \searrow 4$:~\cite{Harada:2003dc} 
$
M_{\rm TD}\searrow   1.5 m_{_{F}} \simeq \sqrt{2} m_{_{F}}\,  (< M_\rho, M_{a_1})\,,
$
which is consistent with Eq.(\ref{PCDCmass}) 
and is  contrasted to the ordinary QCD where the scalar mass is larger than those of the
 vector mesons (``higgsless'') within the same framework of ladder
 SD/BS equation approach. Note that  in this calculation 
$M_{\rm TD} /F_\pi \to {\rm const.}\ne 0$ and hence  there is no isolated massless scalar bound states even in the limit $N_f \to N_f^{\rm cr}$. The result would imply a typical value
 \beq
M_{\rm TD}
\simeq 4 F_\pi \simeq 500 \, {\rm GeV}\quad (M_\rho\simeq M_{a_1}\simeq 12 F_\pi \simeq 1.5 \, {\rm GeV})\,,
\eeq
 in the case of the one-family TC model with 
$F_\pi \simeq 125 \,{\rm GeV}$.

\subsection{Holographic Techni-dilaton~\cite{Haba:2010hu}}
The mass of TD was also calculated 
in a hard-wall-type 
bottom-up holographic TC with $\gamma_m =1$~\cite{Haba:2008nz}  
by including effects of (techni-) gluon 
condensation 
 through the bulk flavor/chiral-singlet scalar field $\Phi_X$, 
in addition to the conventional bulk scalar field $\Phi$ dual to the chiral condensate. 
The TD, a flavor-singlet scalar bound state of
techni-fermion and anti-techni-fermion, will be identified with the lowest KK mode 
coming from the bulk scalar field $\Phi$, not $\Phi_X$.   
Our model with $\gamma_m =0$ and $N_f=3$ 
well reproduces the real-life QCD. 

We consider a couple of typical models of  WTC  
with $\gamma_m \simeq1$ and $N_{\rm TC} = 2,3,4$ based on the CBZ-IRFP  
in the large $N_f$ QCD. The TD mass can be calculated in terms of S parameter and the techni-gluon condensate
normalized by the QCD value: 
$
\Gamma 
\equiv 
\left(
\frac{1}{\pi}\GB/F_\pi^4
\right)^{1/4} \Big|_{normalized}.
$
 For a fixed $S$ the TD mass decreases as $\Gamma$ increases: $M_{\rm TD}/m_{_{F}} \searrow 0$ as $\Gamma \nearrow\infty$.  Using 
the non-perturbative scale  
anomaly Eq.(\ref{npanomaly}) together with the non-perturbative beta function Eq.(\ref{Miranskybeta}),  
we can estimate $\Gamma$ in terms of  
$\Lambda_{\rm ETC}/
m_{_{F}} $:  $\Gamma \nearrow\infty$ as $\Lambda_{\rm ETC}/m_{_{F}} \nearrow\infty $ ($\beta^{^{\rm NP}} (\alpha) \searrow 0$).
In the case of $N_{\rm TC}=3$ ($N_f = 4 N_{\rm TC}, N_D=4$) and $S (\simeq N_D \cdot 8\pi (F_\pi/M_\rho)^2\simeq 8\pi (v/M_\rho)^2)=0.1$,
we have $\Gamma \simeq 7$ for 
$\Lambda_{\rm ETC}/m_{_{F}}  
= 10^3$--$10^4$
(FCNC constraint), which yields
\beq
M_{\rm TD} \simeq 600 \, \GeV\,,
\label{prediction}
\eeq
relatively light 
compared with $M_\rho \simeq M_{a_1} \simeq  3.8 \, \TeV (\simeq v \sqrt{8\pi/S}) $.~\footnote{Note that largeness of $M_\rho$ and $M_{a_1}$ is essentially determined 
by the requirement of $S = 0.1$ fairly independently of techni-gluon 
condensation (Ladder result $M_\rho \simeq 1.5 {\rm TeV}$ corresponds to $S \simeq 0.056N_{\rm TC}N_D\simeq 0.67$ for $N_{\rm TC}=3, N_D=4$). The calculated  $S$ parameter here was from the TC dynamics alone and the actual $S$ parameter could be drastically 
changed by the ETC-like dynamics. For instance, the fermion delocalization~\cite{Cacciapaglia:2004rb} in the Higgsless models as a possible analogue of certain ETC effects in fact  can cancel
large positive $S$ arising from  the 5-dimensional gauge sector which corresponds to the pure TC dynamics.  If it is the case in the explicit ETC model, then the large $S$ value from WTC sector would still be viable and the overall mass scale of techni-hadrons (and hence TD mass also) would be much lower than the above estimate. 
}

\subsection{Discovering Walking Technicolor at LHC~\cite{Matsuzaki:2011ie}}
Now we come to discussions on the discovery signatures of TD at the ongoing LHC. 
Here we take the TD mass as a free parameter in the region $200 \,{\rm GeV}<M_{\rm TD} <1000\,{\rm GeV}$,
since the explicit calculations so far  based on the ladder-like approximation $M_{\rm TD}\simeq 500-600 \,{\rm GeV}$
may have large uncertainty.

\begin{wrapfigure}{r}{6.6cm}
\vspace{-0.5cm}
\includegraphics[width=5.0cm]{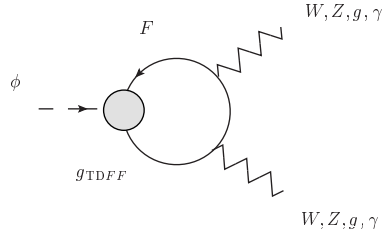}
\caption{ 
The TD couplings to $WW,ZZ,gg,\gamma\gamma$ induced from techni-fermion loops. 
\label{gauge:fig}
}
\end{wrapfigure}

The coupling of TD ($\phi$)  to the SM particles are all through the loop of the techni-fermions ($F$) via the Yukawa coupling of the TD to the techni-fermions, since there is no direct coupling of TD to the SM particles. The Yukawa coupling to the techni-fermions 
 and to the quarks/leptons ($f$) were obtained long time ago~\cite{Bando:1986bg}:
\beq
g_{_{{\rm TD} \, \bar F F}} =
\frac{m_{_{F}}}{v_{\rm TD}}
\,,\quad
g_{_{{\rm TD} \, \bar f f}} =
\frac{m_f}{v_{\rm TD}}
\,,\quad \left(v_{\rm TD} \equiv \frac{F_{\rm TD}}{3-\gamma_m}\simeq \frac{F_{\rm TD}}{2}\right)
\eeq 
where $3-\gamma_m \simeq 2$ for the WTC, and the TD decay constant $F_{\rm TD}$  differs from that of the SM Higgs $v=246 \,{\rm GeV}$.
The coupling to the SM gauge bosons
  ${\cal L}_{{\rm TD}\,WW/ZZ} 
=
g_{{\rm TD}\, WW} \, \phi W_{\mu}^+ W^{\mu -} 
+ \frac{1}{2} (W \rightarrow Z)   
$, 
$  {\cal L}_{{\rm TD}\, gg/\gamma\gamma} 
  = - g_{{\rm TD}\,gg} \,  \phi {\rm tr}[G_{\mu\nu}^2] 
   - g_{{\rm TD}\, \gamma\gamma} \,  \phi F_{\mu\nu}^2 
$ 
are  given by Fig. \ref{gauge:fig}~\cite{Matsuzaki:2011ie}.  The $g_{_{{\rm TD}\,W/Z}}$ 
was evaluated by the direct graphical computation~\cite{Hashimoto:2011cw}, while all the couplings were systematically obtained by the nonlinear realization of the scale and electroweak symmetries
incorporating the explicit breaking of the scale symmetry (due to the dynamical mass generation itself)
via spurion method~\cite{Matsuzaki:2011ie,MY} :
\beq
g_{_{{\rm TD}\,W/Z}} =
\frac{
2m_{W/Z}^2}{v_{\rm TD}},\,\, \,   g_{{\rm TD} \,gg} = 
  \frac{1
  }{v_{\rm TD}} \frac{\beta_F(\alpha_s)}{2 \alpha_s} 
  ,\,\,\, 
    g_{{\rm TD} \,\gamma\gamma} =
\frac{ 1
}{v_{\rm TD}} \frac{\beta_F(\alpha_{\rm EM})}{4 \alpha_{\rm EM}}
\,,\label{TD-gg2gamma:2}
\eeq
where $\beta_F$ stands for the one-loop beta function solely due to the techni-fermions (For the actual calculations
we will also include loop
contributions of the $W/Z$ bosons and the top quark, which are higher loop effects numerically non-negligible (not a substantial effects, though).) 
(Somewhat different estimation was made~\cite{Goldberger:2007zk}.)

Thus the TD couplings differ from those of the SM Higgs by the characteristic factor $1/v_{\rm TD}
$ 
versus
$1/v=1/(246 \, {\rm GeV})$, and also by $\beta_F/v_{\rm TD}$ versus $\beta_{\rm SM}/v$, where $\beta_{\rm SM}$
is the one-loop beta function of only the SM particle contributions. Once we fix the TD mass $M_{\rm TD} $, we may estimate $F_{\rm TD}$ (and hence $v_{\rm TD}$) through the Partially Conserved Dilatation Current (PCDC) hypothesis for the non-perturbative anomaly
Eq.(\ref{npanomaly}):
$F^2_{\rm TD}M^2_{\rm TD}  
= -4 \VEV{\theta^{\mu}_{\mu}}^{^{\rm NP}} 
=
4 m_{_{F}}^4\cdot \kappa_V\, \frac{N_f N_{\rm TC}}{2\pi^2}\,.
$ 
Combining this with Eq.(\ref{PS}), we may have the overall ratio of 
the TD coupling to that of the SM Higgs  $g_{\rm TD}/g_H=v/v_{\rm TD}$:\begin{eqnarray} 
\frac{g_{\rm TD}}{g_{H}} =\frac{v}{v_{\rm TD}}
\simeq
\frac{1}{4 \sqrt{2} \pi} \sqrt{\frac{\kappa_F^4}{\kappa_V}} N_D \frac{M_{\rm TD}}{v}
\simeq 1.4 \left(\frac{N_D}{4} \right) \left(\frac{M_{\rm TD}}{600 {\rm GeV}}\right)
\,, 
\label{coupling}
\end{eqnarray}
where use has been made of  $N_f \simeq 4N_{\rm TC}$, 
$\gamma_m \simeq 1$ and $(\kappa_F, \kappa_V) \simeq (1.5, 0.76)$~\cite{Hashimoto:2010nw} . 

\begin{wrapfigure}{c}{13.2
cm}

 \includegraphics[width=6.6cm]{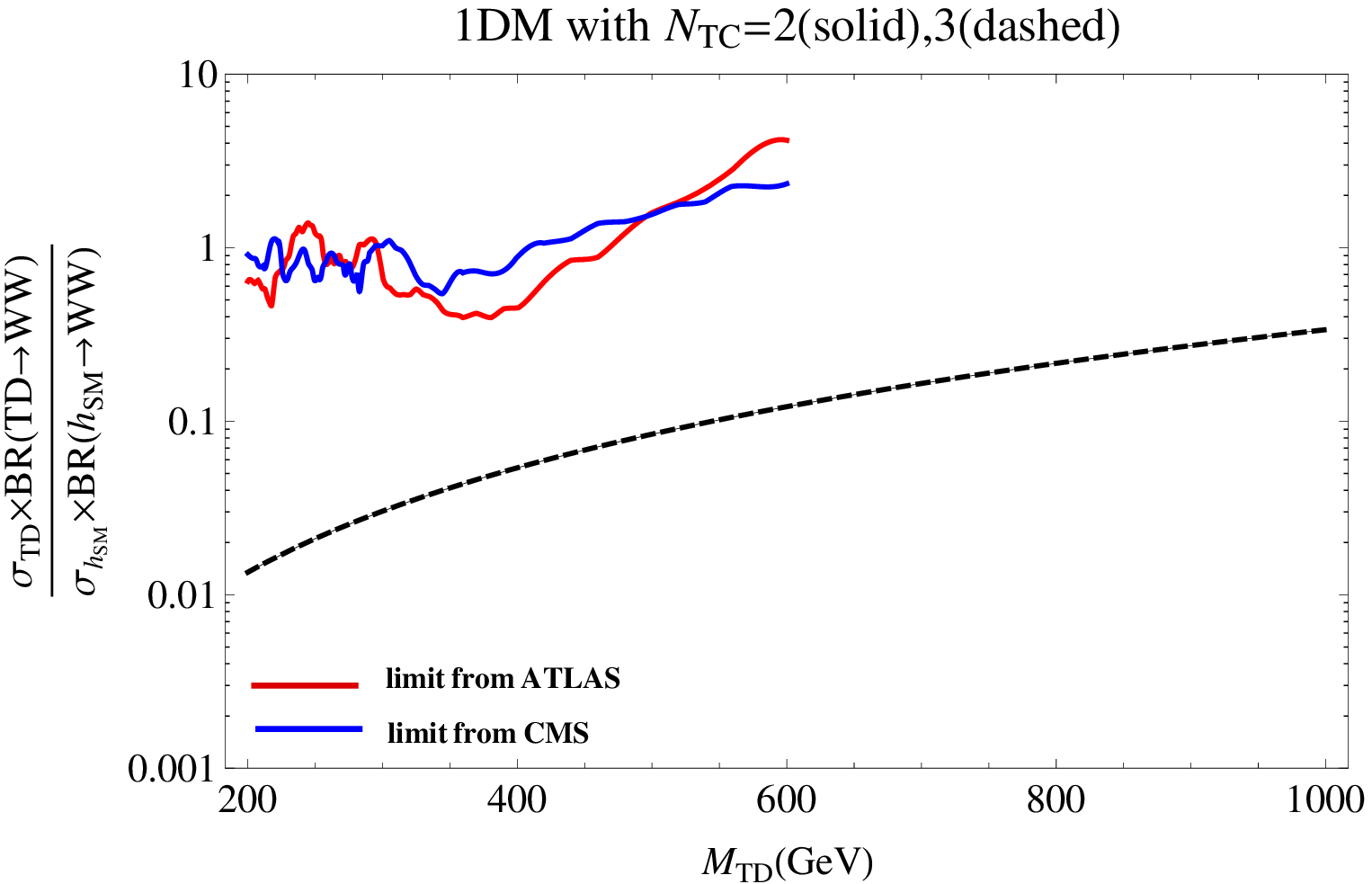}
 \includegraphics[width=6.6cm]{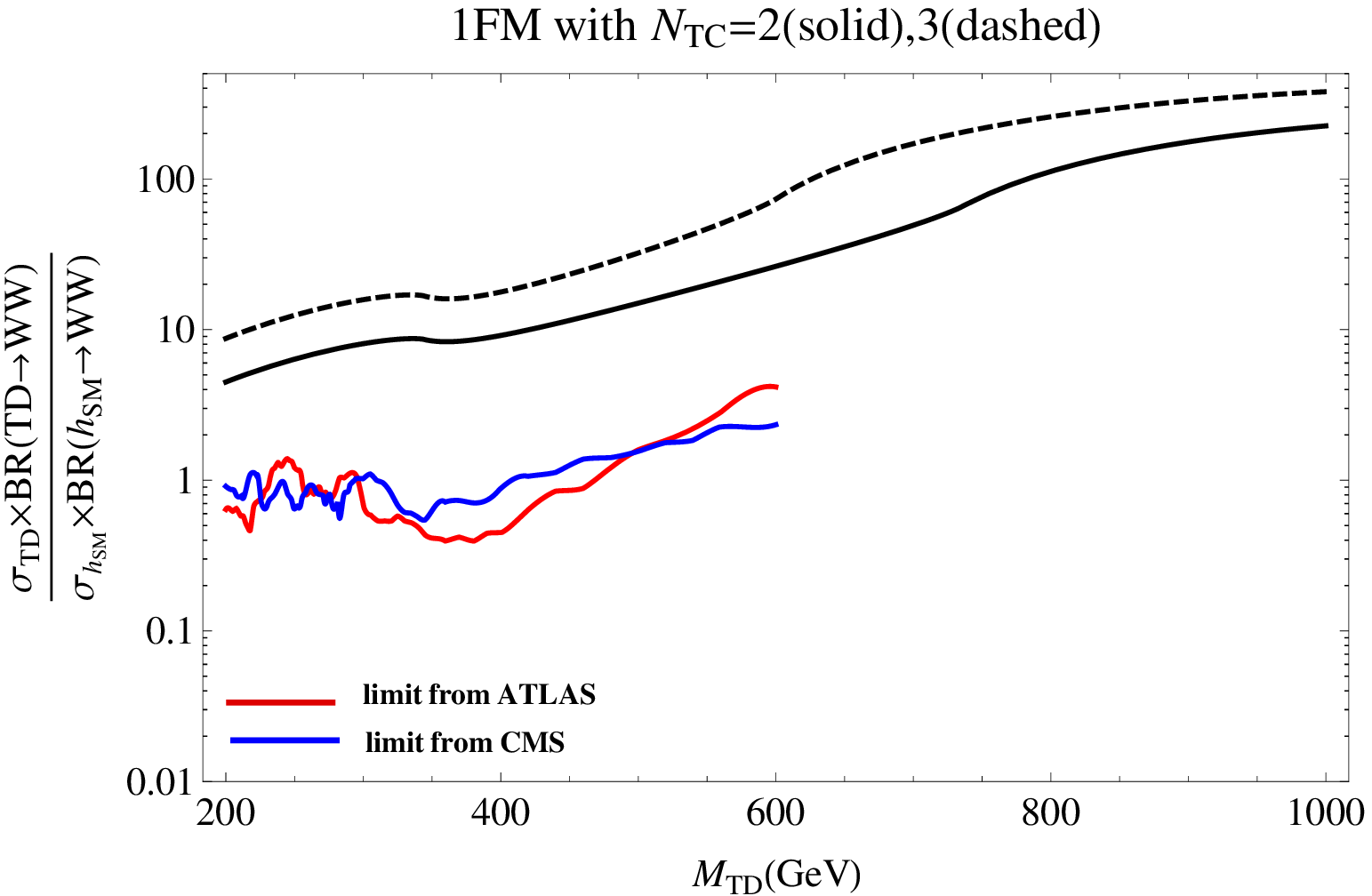}
   \caption{ 
Left panel: The TD LHC production cross sections at $\sqrt{s}=7$ TeV times the $WW/ZZ$ branching ratio 
in the 1DMs with $N_{\rm TC}=2,3$ normalized to the corresponding quantity for the SM Higgs.  
Also shown is the comparison with the 95\% C.L. upper limits from the ATLAS
and CMS.~\cite{ATLAS}
Right 
panel: 
The same as the left panel for the 1FMs.   
\label{LHCsignature1}
}
\end{wrapfigure}

\begin{wrapfigure}{r}{4.5
cm}
\includegraphics[width=4.5cm]{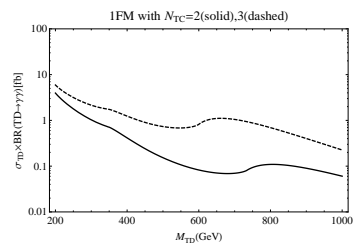}
\caption{ 
The TD  LHC production cross section  at $\sqrt{s}=7$ TeV times the $\gamma\gamma$ branching ratio in unit of fb 
for the 1FMs with $N_{\rm TC}=2,3$.  
\label{LHCsignature2}
}
\end{wrapfigure}

Thus the coupling itself is roughly the same as the SM Higgs.  
The predicted signatures of Higgs-like particles searched at LHC of
the  7 TeV run is depicted in Fig. 
\ref{LHCsignature1} and \ref{LHCsignature2} for the mass range 
$200\,{\rm GeV} <M_{\rm TD} <1000\,{\rm GeV}$.~\cite{Matsuzaki:2011ie}  
In the simplest model, one-doublet model (1DM, $N_D=1$), all the couplings are suppressed relative to the SM Higgs and hence 
invisible at the present LHC search.  In the one-family model (1FM), on the other hand, what is
dramatically different from the SM Higgs  is the 2-gluon and 2-photon couplings which have an extra factor of beta function
$\beta_F$:
There are many techni-fermions having color and electric charges, so that we have a {\it big enhancement 
of gluon fusion production} of TD (by the factor 31(87) for $N_{\rm TC}=2(3)$) in comparison with the SM Higgs, which boosts overall scale of the all decay
channels including a typical searched channel $WW$ at LHC (Fig.\ref{LHCsignature1}).
 Also {\it enhanced are the branching ratios of 2-gluons} (by 16(44) for $N_{\rm TC}=2(3)$) and {\it 2-photon channels} (by 3.2(11) for $N_{\rm TC}=2(3)$) in 1FM relative to the SM Higgs, while other channels are 
roughly the same as in the SM Higgs.
 
Comparing the result with
 the present LHC data~\cite{ATLAS} (see also footnote $\sharp$3) shown in  Fig.\ref{LHCsignature1},
 we see a big chance to {\it discover the 1FM TD for the mass range $M_{\rm TD} > 600 \,{\rm GeV}$}, although
 TD for $200 \,{\rm GeV} < M_{\rm TD} < 600 \,{\rm GeV}$ of the 1FM  is excluded by the present LHC data.   Also we can expect 
in future a {\it prominent signal of 2-photon channel on 0.1-1.0 fb level at $M_{\rm TD} > 600 \,{\rm GeV}$} as shown in Fig.\ref{LHCsignature2}.

\subsection{Discovering Walking Technicolor on the Lattice}
Since the WTC models near conformality are strong coupling theories and the ladder approximation/holographic calculations  would be no more than a qualitative hint,  more reliable calculations   are certainly needed, including the lattice simulations, before drawing a definite conclusion about the physics predictions. Recently there have been many lattice studies of large $N_f$ QCD.~\cite{lattice}  Besides the phase diagram including the TC-induced/ETC-driven four-fermion couplings on the lattice, more reliable calculations of the spectra, particularly the TD,  as well as the anomalous dimension, non-perturbative
beta function, 
$S$ parameter, etc.  are highly desired. For that end we have started to do lattice simulations at KMI (LatKMI collaboration)~\cite{Aoki:2012kr}  on
the large $N_f$ QCD as a possible candidate for the WTC.

\section{Top Quark in Walking Theories}
The top quark is very special in the ETC, since the top mass is too large to be accounted for by the WTC with $\gamma_m \simeq 1$ in the usual ETC-like scenario: 
 $m_{t/b} \sim \frac{1}{\Lambda_{\rm ETC}^2}\, \langle \bar F F \rangle_{\Lambda_{\rm ETC}}\sim m_{_{F}}(m_{_{F}}/\Lambda_{\rm ETC})^{2-\gamma_m}$. Moreover 
it would  require large isospin violation in the condensate  $\langle \bar U U\rangle_{\Lambda_{\rm ETC}} \gg \langle \bar D D\rangle_{\Lambda_{\rm ETC}}$ in order to produce large mass splitting $m_t \gg m_b$, which would in general contradict 
the $T$ parameter constraint, unless we have $\langle \bar U U\rangle_{m_{_{F}}} \simeq \langle \bar D D\rangle_{m_{_{F}}} $.

A possible way-out would be the Strong-ETC TC~\cite{Miransky:1988gk}  
which has an anomalous dimension larger than that of the WTC
 as in the gauged NJL model;  $1<\gamma_m <2$~\cite{Miransky:1988gk} only for $\langle \bar U U\rangle$. This can boost the top mass
 as large as $m_t = {\cal O} (m_{_{F}})$ ($\gamma_m \simeq2$),
  where $m_{_{F}} \simeq 300\, {\rm GeV}\cdot \sqrt{(4/N_D)(3/N_{\rm TC})}$. There wold be no conflict with the
  $T$ parameter as far as we have
  $\langle \bar U U\rangle_{\Lambda_{\rm ETC}} \gg \langle \bar D D\rangle_{\Lambda_{\rm ETC}}$ while keeping
  $\langle \bar U U\rangle_{m_{_{F}}} \simeq \langle \bar D D\rangle_{m_{_{F}}} $. The TD in this case would also have a mass 
   $\sim    500-600 \, {\rm GeV}$ in the ladder approximation, whereas the coupling to the top quark pair will be scaled as $(3-\gamma_m)/F_{\rm TD}\simeq 1/F_{\rm TD}$ instead of  $2/F_{\rm TD}$ of WTC with $\gamma_m\simeq 1$, suppressing the $t \bar t$ channel relative to others  
  by roughly a  factor $1/4$.~\cite{MY}
  
Another possibility to have large top mass is the Top Quark Condensate (Top-Mode Standard Model, TMSM) \cite{Miransky:1988xi} which introduced the $\langle \bar t t \rangle$ based on the dynamics of gauged NJL model
having $\gamma_m \simeq 2$~\cite{Miransky:1988gk} .
The model predicted   (long before the discovery of the top with mass of this large)  the qualitative reality
 that among other quarks
only the top   (as well as $W,Z$) has mass on the order of weak scale. 
  The model also has a composite Higgs boson as a bound state of $\bar t t $ (``Top-Higgs'') 
$m_t  < m_{H_t}  <2m_t$, which
seems to be ruled out by the LHC experiments. A viable Top-mode model involving the top quark condensate would be the top-seesaw model~\cite{Dobrescu:1997nm} and its combination with WTC (``top-seesaw-assisted technicolor'')~\cite{Fukano:2011fp} .

\section{Conclusion}
In the spirit of Sakata we have developed possible composite Higgs models with large anomalous dimension, Walking Technicolor (WTC), Strong
ETC Technicolor, Top-Mode models, etc. 

Particularly the Techni-dilaton (TD) was predicted as a pseudo Nambu-Golodstone boson of the approximate scale symmetry inherent to the WTC. This is a composite Higgs similar to the SM Higgs boson and hence will  be the target of the most urgent Higgs search at LHC. In the typical one-family model having four doublets corresponding to techni-quarks and techni-lepton, mass $M_{\rm TD}$ was estimated to be around 500-600 GeV in the ladder-like computation and/or its combination with the holographic method, although the result should have large uncertainty (The simplest model, one-doublet model, will give a doubled value of mass). Treating $M_{\rm TD}$  as a free parameter, we estimated  the coupling in terms of $M_{\rm TD}$
through PCAC relation, based on the ladder estimation of the vacuum energy and techni-pion decay constant $F_\pi$, see Eq.(\ref{coupling}).
Crucial difference of one-family model from the SM Higgs comes from the enhanced 2-gluon and 2-photon couplings in Eq.(\ref{TD-gg2gamma:2}), which have  big contributions
 from the  loops of the techni-fermions
carrying color and electric charges in the form of the beta functions of color and electromagnetic couplings. One-doublet model lacking these enhancement as well as smaller coupling by factor 1/4 ($=4/N_D$ in Eq.(\ref{coupling})) will give a very small signals and would not be visible at the present setting of LHC.

The actual analysis was performed for the mass range $200 \,{\rm GeV}< M_{\rm TD} < 1,000 \, {\rm GeV}$. See Fig.\ref{LHCsignature1}
and Fig.\ref{LHCsignature2}. The result implies that the one-family TD in the mass region $200 \,{\rm GeV}< M_{\rm TD} < 600 \, {\rm GeV}$ is excluded. There will be a big chance to discover the one-family TD in the region $M_{\rm TD} > 600 \,{\rm GeV}$ because of the large excess
which will be detected in $WW$ channel and 2-photon channel at LHC in near future.

\section*{Acknowledgements}
We thank Shinya Matsuzaki for collaborations for the Techni-dilaton phenomenology which is most relevant to this talk.
This work was supported in part by the JSPS Grant-in-Aid for the Scientific Research  (S) $\sharp$22224003 and 
(C) $\sharp$23540300.

 \end{document}